\documentclass[11pt]{article}
\usepackage{moriond,epsfig}

\def\al{\alpha}

\def\bea{\begin{eqnarray}}
\def\eea{\end{eqnarray}}

\def\beq{\begin{equation}}
\def\eeq{\end{equation}}
\def\beqar{\begin{eqnarray}}
\def\eeqar{\end{eqnarray}}
\def\barr#1{\begin{array}{#1}}
\def\earr{\end{array}}
\def\bfi{\begin{figure}}
\def\efi{\end{figure}}
\def\btab{\begin{table}}
\def\etab{\end{table}}
\def\bce{\begin{center}}
\def\ece{\end{center}}

\def\text{\textstyle}

\def\al{\alpha}
\def\be{\beta}
\def\ga{\gamma}

\def\la{\lambda}
\def\si{\sigma}

\def\De{\Delta}

\def\reffi#1{\mbox{Fig.~\ref{#1}}}

\def\citere#1{\mbox{Ref.~\cite{#1}}}
\def\citeres#1{\mbox{Refs.~\cite{#1}}}

\newcommand{\GeV}{\unskip\,\mathrm{GeV}}

\def\mathswitchr#1{\relax\ifmmode{\mathrm{#1}}\else$\mathrm{#1}$\fi}

\newcommand{\PW}{\mathswitchr W}
\newcommand{\PZ}{\mathswitchr Z}
\newcommand{\PA}{\mathswitchr A}
\newcommand{\Pg}{\mathswitchr g}
\newcommand{\PH}{\mathswitchr H}
\newcommand{\Pe}{\mathswitchr e}

\newcommand{\Ph}{\mathswitchr h}

\newcommand{\Ps}{\mathswitchr s}
\newcommand{\Pb}{\mathswitchr b}

\newcommand{\Pt}{\mathswitchr t}

\def\mathswitch#1{\relax\ifmmode#1\else$#1$\fi}

\newcommand{\MW}{\mathswitch {M_\PW}}

\newcommand{\MZ}{\mathswitch {M_\PZ}}
\newcommand{\MH}{\mathswitch {M_\PH}}

\newcommand{\Mt}{\mathswitch {m_\Pt}}

\newcommand{\mh}{\mathswitch {m_\Ph}}
\newcommand{\mH}{\mathswitch {m_\PH}}
\newcommand{\MA}{\mathswitch {M_\PA}}

\newcommand{\Xt}{X_{\Pt}}

\newcommand{\aeff}{\al_{\mathrm{eff}}}

\newcommand{\mt}{\Mt}

\newcommand{\mgl}{m_{\tilde{\mathrm{g}}}}

\newcommand{\Stop}{\tilde{t}}

\newcommand{\tsf}{\theta\kern-.20em_{\tilde{f}}}
\newcommand{\tsfp}{\theta\kern-.20em_{\tilde{f}\prime}}
\newcommand{\tsq}{\theta\kern-.15em_{\tilde{q}}}

\newcommand{\msusy}{M_{\mathrm{SUSY}}}

\newcommand{\lsim}
{\;\raisebox{-.3em}{$\stackrel{\displaystyle <}{\sim}$}\;}
\newcommand{\gsim}
{\;\raisebox{-.3em}{$\stackrel{\displaystyle >}{\sim}$}\;}

\newcommand{\alps}{\alpha_{\mathrm s}}

\newcommand{\SM}{{\mathrm{SM}}}

\newcommand{\fh}{\textsl{FeynHiggs}}
\newcommand{\subh}{\textsl{subhpole}}

\newcommand{\cp}{{\cal CP}}

\newcommand{\VL}{\left( \begin{array}{c}}
\newcommand{\VR}{\end{array} \right)}
\newcommand{\ML}{\left( \begin{array}{cc}}
\newcommand{\MLd}{\left( \begin{array}{ccc}}
\newcommand{\MLv}{\left( \begin{array}{cccc}}
\newcommand{\MR}{\end{array} \right)}

\newcommand{\gev}{\,\, \mathrm{GeV}}

\newcommand{\BC}{\begin{center}}
\newcommand{\EC}{\end{center}}
\newcommand{\BE}{\begin{equation}}
\newcommand{\EE}{\end{equation}}
\newcommand{\BEA}{\begin{eqnarray}}
\newcommand{\BEAnn}{\begin{eqnarray*}}
\newcommand{\EEA}{\end{eqnarray}}
\newcommand{\EEAnn}{\end{eqnarray*}}

\newcommand{\id}{{\rm 1\kern-.12em
\rule{0.3pt}{1.5ex}\raisebox{0.0ex}{\rule{0.1em}{0.3pt}}}}

\def\draftdate{\relax}
\def\mda{\relax}
\def\mua{\relax}
\def\mla{\relax}
\def\draft{
\def\thtystars{******************************}
\def\sixtystars{\thtystars\thtystars}
\typeout{}
\typeout{\sixtystars**}
\typeout{* Draft mode!
         For final version remove \protect\draft\space in source file
*}
\typeout{\sixtystars**}
\typeout{}
\def\draftdate{\today}
\def\mua{\marginpar[\boldmath\hfil$\uparrow$]%
                   {\boldmath$\uparrow$\hfil}%
                    \typeout{marginpar: $\uparrow$}\ignorespaces}
\def\mda{\marginpar[\boldmath\hfil$\downarrow$]%
                   {\boldmath$\downarrow$\hfil}%
                    \typeout{marginpar: $\downarrow$}\ignorespaces}
\def\mla{\marginpar[\boldmath\hfil$\rightarrow$]%
                   {\boldmath$\leftarrow $\hfil}%
                    \typeout{marginpar:
$\leftrightarrow$}\ignorespaces}
\def\Mua{\marginpar[\boldmath\hfil$\Uparrow$]%
                   {\boldmath$\Uparrow$\hfil}%
                    \typeout{marginpar: $\Uparrow$}\ignorespaces}
\def\Mda{\marginpar[\boldmath\hfil$\Downarrow$]%
                   {\boldmath$\Downarrow$\hfil}%
                    \typeout{marginpar: $\Downarrow$}\ignorespaces}
\def\Mla{\marginpar[\boldmath\hfil$\Rightarrow$]%
                   {\boldmath$\Leftarrow $\hfil}%
                    \typeout{marginpar:
$\Leftrightarrow$}\ignorespaces}
\overfullrule 5pt
\oddsidemargin -15mm
\marginparwidth 29mm
}

\newcommand{\plB}[3]{{\sl Phys. Lett.} {\bf B #1} (19#2) #3}

\newcommand{\prD}[3]{{\sl Phys. Rev.} {\bf D #1} (19#2) #3}

\newcommand{\Cpc}[3]{{\sl Comput. Phys. Commun.} {\bf #1} (20#2) #3}

\newcommand{\NpB}[3]{{\sl Nucl. Phys.} {\bf B #1} (20#2)~#3}
\newcommand{\Nphbps}[3]{{\sl Nucl. Phys.} {\bf B} {\it (Proc. Suppl.)}
{\bf #1B} (20#2) #3}
\newcommand{\PlB}[3]{{\sl Phys. Lett.} {\bf B #1} (20#2) #3}
\newcommand{\PrD}[3]{{\sl Phys. Rev.} {\bf D #1} (20#2) #3}

\begin{document}
\vspace*{4cm}
\title{
\hfill {\normalsize {\rm CERN-TH/2001-212}}\\
\hfill {\normalsize {\rm DCPT/01/64, IPPP/01/32}}\\
\hfill       \mbox{\normalsize {\rm hep-ph/0108063}}\\[2em] 
THEORETICAL IMPLICATIONS OF THE POSSIBLE OBSERVATION\\
OF HIGGS BOSONS AT LEP}

\author{ GEORG WEIGLEIN }

\address{TH Division, CERN, CH-1211 Geneva 23, Switzerland\\
Institute for Particle Physics Phenomenology, University of Durham,
Durham DH1~3LE, UK}

\maketitle\abstracts{
Theoretical implications of the possible observation of a Higgs boson
with a mass of about 115~GeV at LEP are discussed. Within the
Standard Model a Higgs boson in this mass range agrees well with the
indirect constraints from electroweak precision data. However, it 
would nevertheless point towards physics beyond the Standard Model, in 
particular to Supersymmetric extensions. The interpretation of the LEP
excess as production of the light or the heavy $\cp$-even Higgs boson is
discussed within the unconstrained MSSM and the mSUGRA, GMSB and AMSB
scenarios. Prospects for Higgs physics at future colliders are briefly
summarized.
}

\section{Standard Model}

Within the electroweak Standard Model (SM) the Higgs boson is the last
missing ingredient that has not been experimentally confirmed so far.
Its mass, $\MH$, is a free parameter of the theory and is only bounded
from above by unitarity arguments to be below about 1~TeV. In the final
year of LEP running at an average center-of-mass energy of about 206~GeV,
the combined results of the four LEP experiments established a 95\%
C.L.\ exclusion limit for the SM Higgs boson of $\MH > 113.5$~GeV
(expected: $\MH > 115.3$~GeV). An excess of events at about
$\MH \approx 115$~GeV with a significance of $2.9 \sigma$
(corresponding to the probability for a background fluctuation of 
$4.2 \times 10^{-3}$) was observed, which is 
compatible with the production of a SM Higgs boson in this mass 
range~\cite{LEPHiggsSM1100}.%
\footnote{New combined results were presented at the 2001 Summer
Conferences~\cite{LEPHiggsSM0701} based still on preliminary results of
three collaborations and final results of one 
collaboration~\cite{L3HiggsSM0701}. They yield a 95\% C.L.\ exclusion
limit for the SM Higgs boson of $\MH > 114.1$~GeV (expected: $\MH >
115.4$~GeV) and show an excess of events that
can be interpreted as the production of a SM Higgs of about 115.6~GeV. 
The probability for a background fluctuation generating the observed
effect is 3.4\%, corresponding to a significance of $2.1 \sigma$.
}

\begin{figure}
\centerline{
\psfig{figure=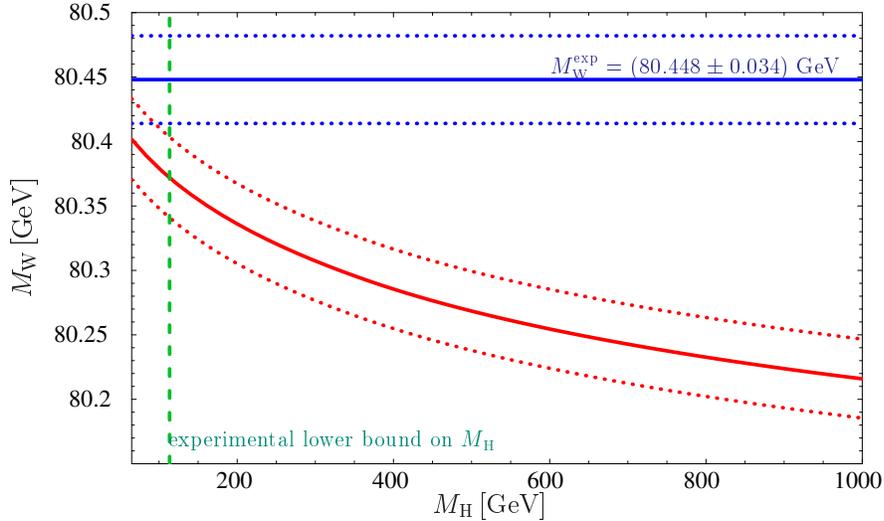,width=12cm} 
}
\caption{The prediction for $\MW$ as a function of $\MH$ within the SM.
It is compared with the experimental value of $\MW$, $\MW^{\rm exp}$,
and the experimental
95\% C.L.\ lower bound on the Higgs boson mass.
\label{fig:mw2l}
}
\end{figure}

Indirect constraints on the Higgs boson mass in the SM can be obtained
by comparing the electroweak precision data with the predictions of the
theory. As an example, \reffi{fig:mw2l} shows the SM prediction for
$\MW$ as a function of $\MH$ based on the result of \citere{delr}
incorporating the complete fermionic contributions at the two-loop level.
The theory
predictions are affected by two kinds of uncertainties: from unknown
higher-order corrections and from the experimental errors of the input
parameters. These uncertainties are indicated in \reffi{fig:mw2l} as a
band of two dashed lines around the central value (given by the solid
line). At present the theoretical uncertainties are dominated by the 
error of the top-quark mass, $\mt = 174.3 \pm 5.1$~GeV, which gives rise
to an uncertainty of $\MW$ of about $\pm 30$~MeV. The prediction for $\MW$ 
as function of $\MH$ is compared in \reffi{fig:mw2l} with the
experimental value, 
$\MW^{\mathrm{exp}} = 80.448 \pm 0.034$~GeV~\cite{mori01data}.
The figure clearly shows a
preference for a light Higgs boson within the SM. In fact, taking into
account the experimental 95\% C.L.\ lower bound on the Higgs boson mass,
$\MH = 113.5$~GeV~\cite{LEPHiggsSM1100}, the allowed intervals of the 
theory prediction and the experimental result have no overlap (at the 
$1 \sigma$ level). The best description of the data within the SM is thus
obtained for a Higgs boson being `just around the corner'
\footnote{The slight discrepancy between the theory prediction for $\MW$
and the experimental value has further increased in the results of
the 2001 Summer Conferences with the experimental value $\MW^{\rm exp} =
80.451 \pm 0.033$~GeV~\cite{summer01data} and the Higgs exclusion bound
of $\MH > 114.1$~GeV~\cite{LEPHiggsSM0701}.
}
(see also \citere{agosterl}).
A global fit to all data yields for the Higgs boson mass within the SM
$\MH = 98^{+58}_{-38}$~GeV, corresponding to a 95\% C.L.\ upper limit of
$\MH < 212$~GeV.%
\footnote{The corresponding results of the 2001 Summer Conferences are
$\MH = 88^{+53}_{-35}$~GeV and $\MH < 196$~GeV at 95\%
C.L.~\cite{summer01data}.
}

While a Higgs boson mass of about 115~GeV would fit well in the
context of the SM from the point of view of electroweak precision data,
a value of $\MH$ in this region would on the other hand be problematic
concerning the stability of the electroweak vacuum.
For $\MH \approx 115$~GeV (and $\mt = 175$~GeV, $\alps(\MZ) =
0.118$) one would expect that new physics is
required at a scale $\Lambda \lsim 10^6 \gev$ in order to prevent the
effective Higgs potential from being destabilized by top-quark loop
corrections~\cite{vacstabil}. It has been argued in 
\citere{ellisross} that the kind of new physics suitable for stabilizing
the electroweak vacuum must share several important features with
Supersymmetric theories. It would require in particular extra bosonic
degrees of freedom and a high degree of fine-tuning of the model
couplings, which is automatically fulfilled in a Supersymmetric theory.

\begin{figure}
\centerline{
\psfig{figure=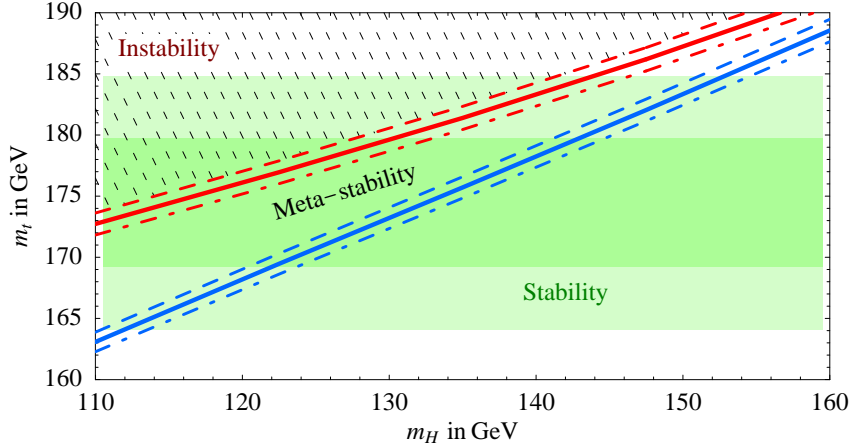,width=12cm} 
}
\caption{Regions in the $(\MH, \mt)$ plane with stability,
meta-stability and instability of the Standard Model vacuum 
if no new physics below the Planck scale is present. The solid lines
refer to $\alps(\MZ) = 0.118$, while the dashed and dot-dashed lines
correspond to $\alps(\MZ)=0.118\pm 0.002$. The shaded area indicates
the experimental range for $\mt$. Possible effects of subleading
contributions are estimated to shift the bounds by $\pm 2\GeV$ in $\mt$.
\label{fig:metastab}
}
\end{figure}

The above arguments concerning the need for new physics at a 
scale $\Lambda \lsim 10^6 \gev$ cannot be regarded as fully rigorous, 
since they rely on the rather strong requirement that the minimum of the
effective Higgs potential should be absolutely stable (and that $\mt$
should not be smaller than 1--2~$\si$ below its current experimental 
central value). A detailed analysis of the case of a metastable vacuum
shows that for $\MH \approx 115$~GeV and $\mt \approx 175$~GeV
(or smaller values of $\mt$) the electroweak vacuum can be sufficiently 
long lived with respect to the age of the universe even without new physics 
below the Planck scale, see \reffi{fig:metastab}~\cite{metastabil}.
Nevertheless, the arguments above underline that Supersymmetry (SUSY)
provides a very attractive framework for naturally accommodating a light Higgs 
boson (and furthermore, independently of arguments relying on the precise value 
of $\MH$, the hierarchy problem points towards new physics at the TeV scale).

\section{Minimal Supersymmetric Standard Model (MSSM)}

In contrast to the SM, the mass of the lightest $\cp$-even Higgs boson
in the MSSM, $\mh$, is not a free parameter but can be predicted from
the other parameters of the model. This gives rise to the upper bound
$\mh < \MZ$ at lowest order. This bound is affected by large
higher-order corrections~\cite{mhcorr,subh,hhh,mhFD,mhiggslong,maulpaulslav}, 
shifting it upwards to about 
$\mh \lsim 135$~GeV at the two-loop level in the unconstrained 
MSSM~\cite{mhiggslong}. This bound stays unaffected if non-zero 
$\cp$-violating phases are allowed~\cite{cphiggsA}.
Since $\mh$ is predicted within the MSSM, the measurement of the
Higgs boson mass provides a more direct test of the model than in the
case of the SM.

For the analysis of the LEP data the theoretical predictions implemented
in the programs \fh~\cite{feh}, based on a Feynman-diagrammatic
two-loop result~\cite{mhFD} (incorporating the complete one-loop
result~\cite{mhonel}), and \subh~\cite{subh}, based on a
renormalization-group improved one-loop effective potential
result~\cite{subh,bse}, are used. The remaining theoretical uncertainties
from unknown higher-order corrections have been estimated to be about
$\De\mh \approx \pm 3$~GeV~\cite{flablcws}. The biggest theoretical
uncertainty at present arises from the experimental error of the
top-quark mass. The current error of about $\pm 5$~GeV in $\mt$ induces
an uncertainty of also about $\pm 5$~GeV in $\mh$~\cite{tbexcl}. Thus,
an accurate measurement of the top-quark mass is mandatory in order to
allow precise theoretical predictions in the MSSM Higgs sector.

Confronting the upper bound on $\mh$ as function of $\tan\be$, the ratio
of the vacuum expectation values of the two Higgs doublets, with the
exclusion bounds obtained at LEP, experimental constraints on $\tan\be$
can be derived. In the $\mh^{\rm max}$ benchmark
scenario~\cite{bench}, which yields the maximum values for
$\mh(\tan\be)$ for fixed $\mt = 174.3$~GeV and $\msusy = 1$~TeV in the
unconstrained MSSM, the
$\tan\be$ region $0.5 < \tan\be < 2.4$ can be excluded~\cite{LEPHiggsMSSM}. 
In the no-mixing benchmark scenario~\cite{bench}, which uses the same
parameters as the $\mh^{\rm max}$ scenario except that vanishing mixing
in the scalar top sector is assumed, only relatively small parameter
regions remain unexcluded, and the region $0.7 < \tan\be < 10.5$ is
ruled out~\cite{LEPHiggsMSSM}.

The main production channels for the neutral MSSM Higgs bosons at LEP
are the Higgs\-strahlung process, $\Pe^+\Pe^- \to \Ph\PZ, \PH\PZ$, 
and the associated
production, $\Pe^+\Pe^- \to \Ph\PA, \PH\PA$. 
For the lightest $\cp$-even Higgs boson the
cross section $\si_{\Ph\PZ}$ is approximately given by $\si_{\Ph\PZ}
\approx \sin^2(\be - \aeff) \si_{\Ph\PZ}^{\SM}$, where
$\si_{\Ph\PZ}^{\SM}$ is the SM cross section for the same Higgs mass
and $\aeff$ is the effective mixing angle of the neutral $\cp$-even 
Higgs bosons.%
\footnote{This approximation is applicable at LEP energies, while 
sizable corrections to it occur at higher energies~\cite{eezha}.}
The cross section for the associated production contains a complementary
factor, $\si_{\Ph\PA} \approx \la \cos^2(\be - \aeff)
\si_{\Ph\PZ}^{\SM}$, where $\la$ is a kinematic factor. 

\begin{figure}[ht]
\centerline{
\psfig{figure=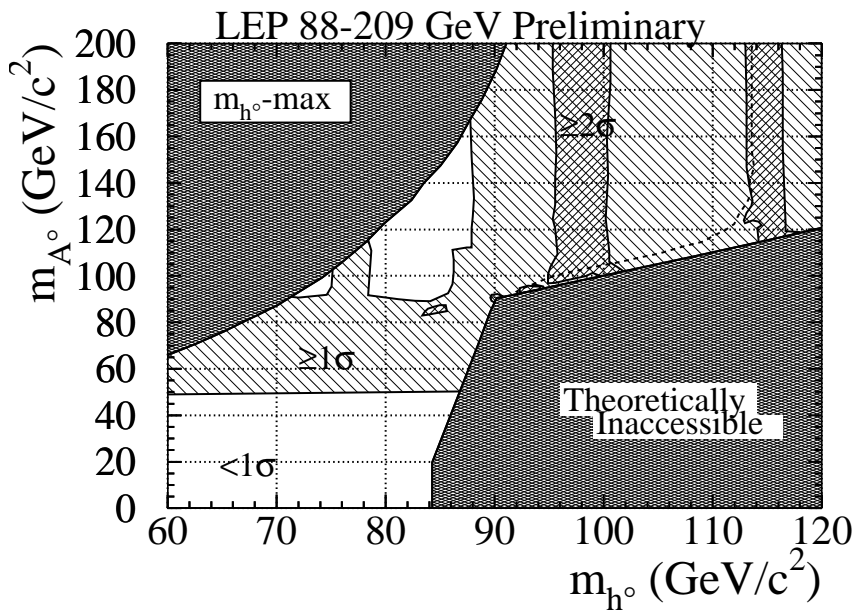,width=12cm,height=9cm}
}
\caption{Excluded region in the $(\mh, \MA)$ plane in the $\mh^{\rm max}$ 
scenario (above and left to the dashed line) and parameter regions where
the observation is more than $1\si$ or more than $2\si$ above the
background prediction. 
\label{fig:mssmhiggs}
}
\end{figure}

The excess in
the search for the SM Higgs boson manifests itself also in the MSSM
Higgs boson searches, see \reffi{fig:mssmhiggs}~\cite{LEPHiggsMSSM}.
The figure shows the 95\% C.L.\ exclusion contour in the $(\mh, \MA)$
plane in the $\mh^{\rm max}$ scenario and the parameter regions where
the observation is more than $1\si$ or more than $2\si$ above the
background prediction. Vertical structures in the plot are due to
features in the $\Ph\PZ$ search results, while structure on the $\mh \approx
\MA$ line arises mainly from the $\Ph\PA$ searches. For $\MA \gg \MZ$ 
$\Ph$ has SM-like couplings (which means in particular 
$\sin^2(\be - \aeff) \approx 1$), and the results for the SM search can
directly be taken over for the MSSM case. The corresponding events give
rise to the vertical structure at $\mh \approx 115$~GeV indicating an excess
above $2\si$ in \reffi{fig:mssmhiggs}. The MSSM also allows, however, 
another more speculative interpretation of the LEP excess. In the
parameter region $\mh, \MA \approx 100$~GeV the $\Ph\PZ\PZ$ coupling is
strongly suppressed, $\sin^2(\be - \aeff) \ll 1$, while the heavy
$\cp$-even Higgs boson $\PH$ has SM-like couplings. Thus, it would in
principle be possible that the excess events observed at LEP were caused
by the production of the heavy $\cp$-even Higgs boson with mass 
$\mH \approx 115$~GeV, while the light $\cp$-even Higgs boson has been
produced in the Higgsstrahlung process with such a small rate that it
could not be observed above the background. In this case the associated
production channel, $\Pe^+\Pe^- \to \Ph\PA$, should be open, but it can be
suppressed or even completely closed if the mass sum $\mh + \MA$ is
close to or above the kinematic limit. 

\begin{figure}[ht]
\centerline{
\psfig{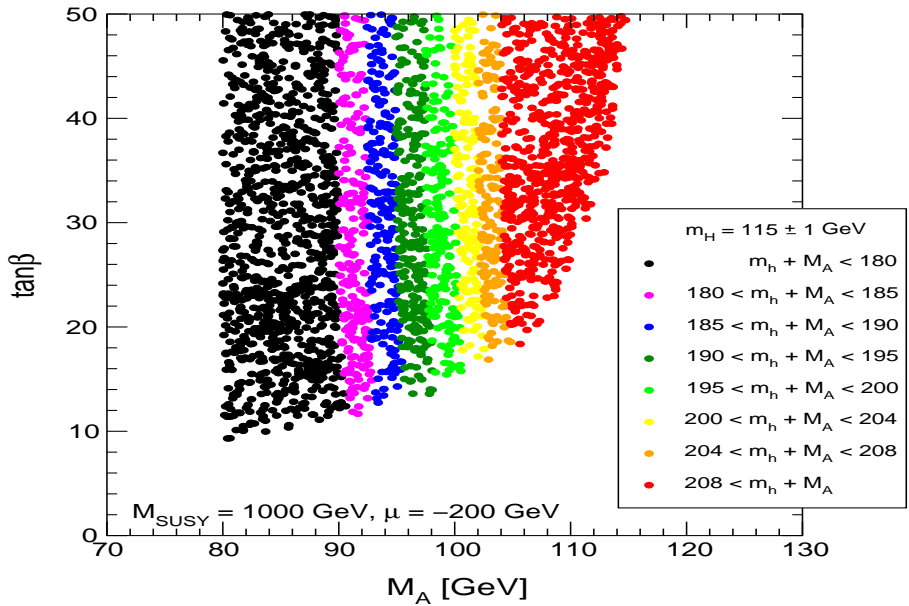}
}
\caption{Allowed parameter space in the $(\MA, \tan\be)$ plane in the
unconstrained MSSM for $\mH \approx 115$~GeV and a significantly
suppressed $\Ph\PZ\PZ$ coupling, $\sin^2(\be - \aeff) < 0.2$. The different 
shadings correspond to different values of $\mh + \MA$.
\label{fig:mssmheavyhiggs}
}
\end{figure}

\reffi{fig:mssmheavyhiggs} shows the allowed parameter space in the 
$(\MA, \tan\be)$ plane in the unconstrained MSSM for $\mH \approx
115$~GeV and a significantly suppressed $\Ph\PZ\PZ$ coupling, $\sin^2(\be -
\aeff) < 0.2$~\cite{mssmheavyhiggs}. The different shadings indicate
different values of $\mh + \MA$. The region $\mh + \MA < 180$~GeV is
excluded by LEP, while the LEP searches have practically no 
sensitivity anymore for $\mh + \MA \gsim 190$~GeV (the
region $\mh + \MA > 208$~GeV is even outside the
kinematic reach of LEP). Accordingly, for the parameter space with 
$\mh + \MA \gsim 190$~GeV in the plot an interpretation of the LEP
excess in terms of the production of the heavy $\cp$-even Higgs boson
appears to be possible. 

\reffi{fig:mssmhiggs} contains another unexcluded region in the 
$(\mh, \MA)$ plane with an excess above $2\si$ for 
$\mh \approx \MA \approx 100$~GeV. The question how well
\reffi{fig:mssmhiggs} is compatible with the production of three MSSM
Higgs bosons at LEP with $\MA \approx 100$~GeV, $\mh \approx 100$~GeV and 
$\mH \approx 115$~GeV has not yet been directly answered 
because the LEP analyses are designed to search for only one $\cp$-even
Higgs boson with a specified mass at a time.%
\footnote{A more detailed investigation of this issue will become
possible on the basis of the result in \citere{LEPHiggsSM0701}, where the 
signal expected from a 115~GeV Higgs was injected in the background simulation
and propagated through the likelihood ratio calculation at each 
$\MH$ value.}
There is a substantial
dilution of the significance of a combination of the two excesses because
statistical fluctuations can occur anywhere in the two-dimensional space
of $\mh$ and $\mH$.


It should
furthermore be noted that qualitatively the same behavior as described
above for the $\cp$-conserving MSSM can happen in an even wider
parameter space if $\cp$-violating phases are allowed. In this case a
strong suppression of the coupling of the lightest Higgs boson to the 
Z~boson can occur, while the next-to-lightest Higgs boson couples to the 
Z~boson with almost SM strength~\cite{cphiggsB}.

\begin{figure}[ht]
\centerline{
\psfig{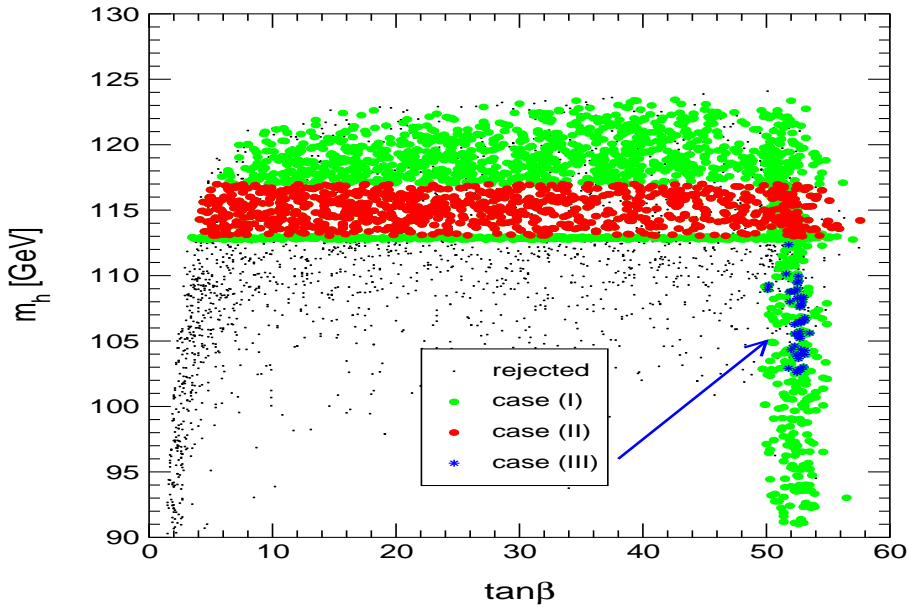}
}
\caption{The light $\cp$-even Higgs boson mass $\mh$ as a function of
$\tan\be$ in the mSUGRA scenario. Allowed parameter points are
indicated by big green points (light shaded, ``case~(I)''), big red 
points (dark shaded,
``case~(II)'') and blue stars (indicated by an arrow in the plot, 
``case~(III)''), while the little black
dots indicate parameter points which are in principle possible in the
mSUGRA scenario but are rejected because of the LEP Higgs bounds and 
further experimental and theoretical constraints. Case~(II) is the
subset of allowed parameter points which are consistent with the 
interpretation of the LEP excess as production of the lightest
$\cp$-even Higgs boson (i.e.\ $\mh \approx 115$~GeV and SM-like
couplings of the $\Ph$), while case~(III) corresponds to the
interpretation of the LEP excess in terms of the heavier $\cp$-even
Higgs boson.
\label{fig:mhSUGRA}
}
\end{figure}

The excess of events observed in the Higgs search at LEP has been
analyzed within different SUSY scenarios by many 
authors~\cite{mh115SUSY,asbs}. For example, in \citere{asbs} a 
comparison of the
mSUGRA, GMSB and AMSB soft SUSY-breaking scenarios has been performed. 
The interpretation of the LEP excess as the production of the lightest
$\cp$-even Higgs boson is possible in all three scenarios, while the 
interpretation in terms of the production of the heavier $\cp$-even
Higgs boson is only possible within the mSUGRA scenario. In
\reffi{fig:mhSUGRA} the allowed parameter space in the $(\tan\be, \mh)$ 
plane is displayed in the mSUGRA scenario, and the regions corresponding
to the two interpretations of the LEP excess are indicated. The LEP
Higgs searches exclude all parameter points with $\mh \lsim 113$~GeV
and $\tan\be \lsim 50$. This is contrary to the situation in the 
$\mh^{\rm max}$ scenario of the unconstrained MSSM, where the exclusion
bound on the SM Higgs boson applies to $\mh$ only for $\tan\be \lsim 8$.
For larger values of $\tan\be$ and small $\MA$, in the unconstrained MSSM
a suppression of the $\Ph\PZ\PZ$ coupling is possible, resulting in a reduced
production rate compared to the SM case. In the mSUGRA scenario, a
significant suppression of the $\Ph\PZ\PZ$ coupling occurs only in a small
allowed parameter region where $50 \lsim \tan\be \lsim 55$. In this
region an interpretation of the LEP excess in terms of the production of
the heavier $\cp$-even Higgs boson is possible (for $\mH \approx 115$~GeV).
It should be noted, however, that this parameter region is close to the
exclusion bounds obtained at Run~I of the Tevatron~\cite{cdfhiggs} and
will soon be probed with the Run~II data~\cite{tevhiggs}.

{}From \reffi{fig:mhSUGRA} one can read off an upper bound on $\mh$ of
$\mh \lsim 124$~GeV in the mSUGRA scenario for $\mt = 175$~GeV, which is
about 6~GeV lower than in the unconstrained MSSM. The lower bound on 
$\tan\be$ in the mSUGRA scenario is $\tan\be \gsim 3.3$, i.e.\ slightly
higher than in the unconstrained MSSM. In the GMSB and AMSB scenarios
the bounds are $\mh \lsim 119$~GeV, $\tan\be \gsim 4.6$ (GMSB) and 
$\mh \lsim 122$~GeV, $\tan\be \gsim 3.2$ (AMSB) for 
$\mt = 175$~GeV~\cite{asbs}.

\section{Prospects for the future}

In Run~II of the Tevatron the main Higgs production channels are the
Higgs\-strahlung from the W and the Z~boson, i.e.\ a similar production
mechanism as at LEP. A Higgs boson with a mass of about 115~GeV, i.e.\
close to the present exclusion bound, would be favorable for the Higgs
search at the Tevatron, and the sensitivity for a 95\% exclusion limit
on the SM Higgs boson could be reached with an integrated luminosity of
about 2~fb$^{-1}$ per experiment~\cite{tevhiggs} (which could be achieved 
in 2003), while the sensitivity for a $5\si$ 
discovery of a SM Higgs boson with $\MH \approx 115$~GeV could be reached 
with about 15~fb$^{-1}$ per experiment~\cite{tevhiggs} (possibly in 2007). 
At the LHC, on the other hand, the sensitivity for a $5\si$ discovery of
a SM Higgs boson in the whole mass range up to 1~TeV can be obtained with 
about 10~fb$^{-1}$~\cite{lhchiggs}. The mass region of about 115~GeV
is the most difficult one for the LHC, where the search relies on the
$\Pg\Pg \to \PH \to \ga\ga$ and $\Pt \bar \Pt \PH$, $\PH \to \Pb \bar \Pb$ 
channels (further sensitivity can be added via the weak boson fusion
channel~\cite{wbf}, which is currently under study). If both machines run on 
schedule, they could reach the sensitivity 
for a $5\si$ discovery of a SM Higgs boson with $\MH \approx 115$~GeV
approximately at the same time.

\begin{figure}[ht]
\centerline{
\psfig{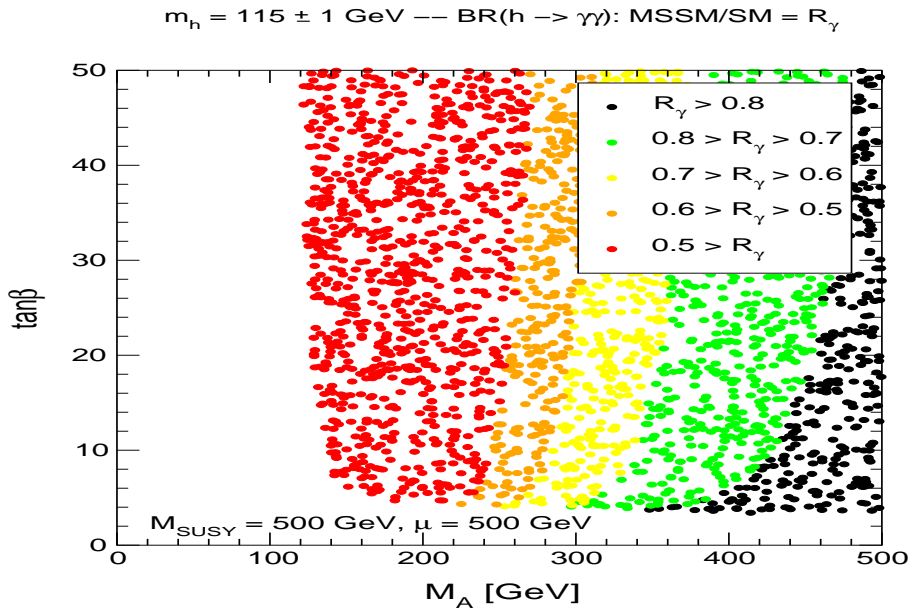}
}
\caption{The branching ratio BR$(\Ph \to \ga\ga)$ normalized to the SM
value, ${\rm R}_{\ga}$, in the 
$(\MA,\tan\be)$ plane in the unconstrained MSSM for $\mh \approx 115$~GeV. 
\label{fig:higgssuppr}
}
\end{figure}

In the MSSM, parameter regions exist where the discovery of a Higgs boson
with about 115~GeV is more difficult than in the SM (see 
e.g.~\citeres{higgsdetect}). As an example, \reffi{fig:higgssuppr}
shows the branching ratio BR$(\Ph \to \ga\ga)$ in the MSSM, normalized
to the SM value, in the unconstrained MSSM for 
$\mh \approx 115$~GeV~\cite{brhagaga}.
As can be seen in the plot, a significant suppression is possible over a
wide parameter range, making this channel less sensitive than in the SM
case.

\begin{figure}[ht]
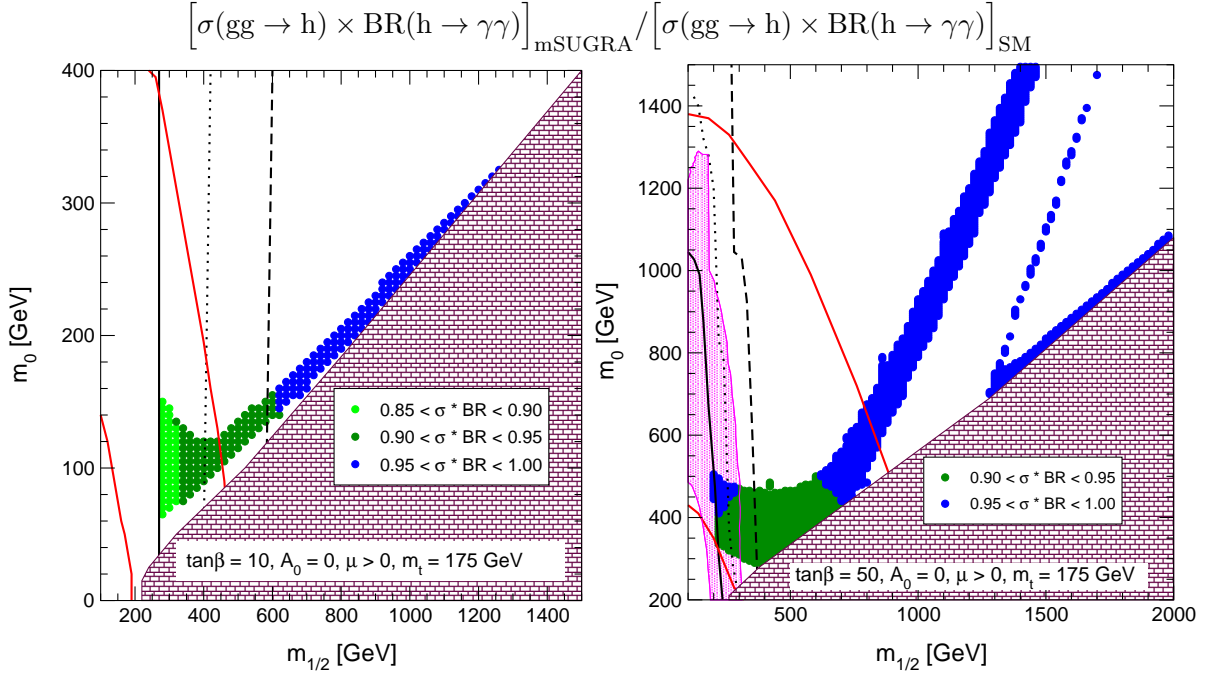


\begin{center}
$\Bigl[\sigma(\Pg\Pg \to \Ph) \times 
       {\rm BR}( \Ph \to \gamma \gamma)\Bigr]_{\rm
mSUGRA}
 / \Bigl[\sigma(\Pg\Pg \to \Ph) \times {\rm BR}( \Ph \to \gamma
\gamma)\Bigr]_{\rm
SM} $
\end{center}

\begin{minipage}{8in}
\epsfig{file=EHOW03c.03.cl.eps,height=3.2in}
\epsfig{file=EHOW03c.09.cl.eps,height=3.2in} \hfill
\end{minipage}

\caption{The cross section for production of the lightest $\cp$-even
Higgs boson in gluon fusion and its decay into a photon pair,
normalized to the SM value with the same Higgs mass, is shown in the
$(m_{1/2}, m_0)$ planes of the mSUGRA scenario for the cosmologically
allowed parameter region.
The diagonal (red) solid lines are the $\pm 2 \sigma$ contours for 
$g_\mu - 2$.
The near-vertical solid, dotted and dashed (black) lines are the $\mh =
113, 115, 117$~GeV contours. The light shaded (pink) regions are excluded 
by $\Pb \rightarrow \Ps \gamma$. The (brown) bricked
regions are excluded since in these regions the lightest SUSY particle
is the charged $\tilde\tau_1$.
\label{fig:higgssugralhc}
}
\end{figure}

The situation is different, however, if one focuses on the mSUGRA
scenario and furthermore takes into account constraints from the
cosmological relic density and the results for $\Pb \to \Ps \ga$ and
$g_{\mu}-2$. \reffi{fig:higgssugralhc} shows the parameter space 
consistent with the dark matter constraint in the
$(m_{1/2}, m_0)$ plane of the mSUGRA scenario for $A_0 = 0$ and $\mu > 0$
for two values of $\tan\be$~\cite{ehow}. The preferred regions from the 
LEP Higgs search, $\Pb \to \Ps \ga$ and $g_{\mu}-2$ are indicated in the plots.
The figure shows that for both values of $\tan\be$ there is a parameter
space within the mSUGRA scenario that is consistent with all
experimental constraints. The different shadings correspond to different
values of $\sigma(\Pg\Pg \to \Ph) \times {\rm BR}( \Ph \to \gamma \gamma)$,
normalized to the SM value. As a result (which is also valid for non-zero
values of $A_0$), no significant suppression of
the $\Pg\Pg \to \Ph \to \gamma \gamma$ production channel at the LHC
occurs, and the lightest $\cp$-even Higgs boson should be discoverable
at the LHC with 10~fb$^{-1}$ in this scenario. Similar results hold for
the $\Pt \bar \Pt\Ph$ associated production at the LHC and the Higgsstrahlung
processes at the Tevatron~\cite{ehow}.

At a future Linear Collider (LC) precision measurements of the Higgs
mass and its couplings to gauge bosons and fermions (including the 
$\Pt \bar \Pt \PH$ coupling) will become possible~\cite{lcreps}. 
The LC measurements
will furthermore provide informations on the triple Higgs self-coupling,
which will be important for reconstructing the Higgs potential. They
will furthermore allow to determine the spin and parity quantum numbers
of the Higgs boson.
Thus, the LC measurements will be important in order to experimentally
establish the Higgs mechanism. Studying the recoil against the Z~boson,
at the LC the production via Higgsstrahlung can be studied completely
independent from the Higgs decay modes, which is in particular important
if the Higgs boson has a large branching fraction into invisible decay
products.
Furthermore, a precise measurement of the top-quark mass with an
accuracy of $\De\mt^{\rm exp} \lsim 200$~MeV at the LC will be
indispensable in order to match the experimental precision of the $\mh$
measurement at the LHC with the accuracy of the theoretical prediction
within the MSSM.

\begin{figure}[ht]
\centerline{
\psfig{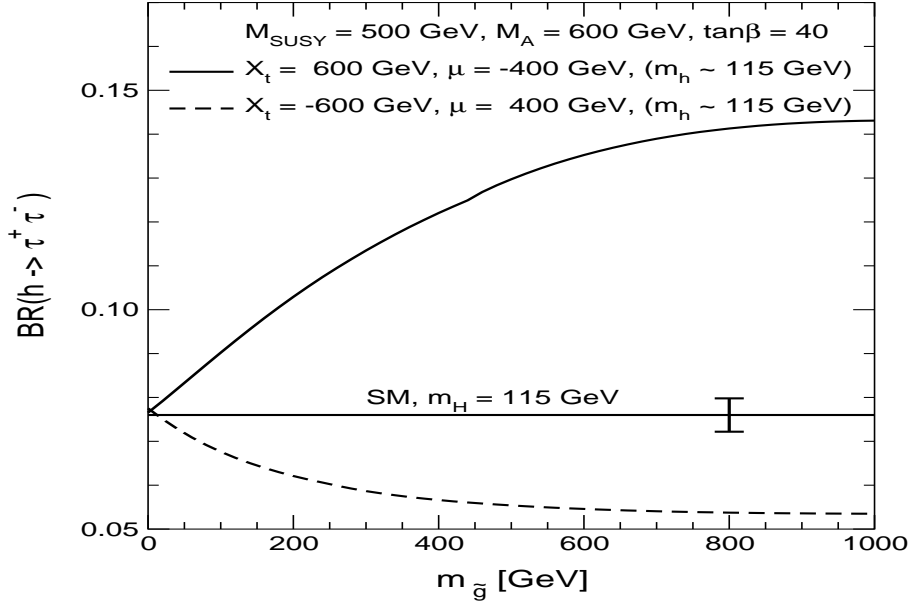}
}
\caption{Prediction for BR($\Ph \to \tau^+ \tau^-$) in the MSSM
as a function of the gluino mass for $\mh \approx 115$~GeV and 
two values of the off-diagonal entry in the $\Stop$ mixing matrix, 
$\Xt$, and the Higgs mixing parameter, $\mu$. The SM value is also
shown. The error bar at the SM prediction indicates the prospective
experimental accuracy at the LC.
\label{fig:taubr}
}
\end{figure}

In the context of an assumed observation of 
a Higgs boson with a mass of about 115~GeV, the precision
measurements at the LC will allow a very sensitive test of the model.
As an example, \reffi{fig:taubr} shows the prediction for BR($\Ph \to
\tau^+ \tau^-$) in the MSSM as a function of the gluino mass, $\mgl$, in
comparison with the SM prediction and the prospective experimental
accuracy at the LC of about 5\%~\cite{flablcws,hff}. Large gluino
and higgsino loop corrections can affect the $\Ph \Pb \bar \Pb$ coupling 
for large values of $\tan\beta$ and/or $\mu$ and can thus give rise to a
sizable shift in BR($\Ph \to \tau^+ \tau^-$). A precise measurement
of BR($\Ph \to \tau^+ \tau^-$) at a future LC will thus provide a high
sensitivity for a distinction between the SM and the MSSM even for
relatively large values of $\MA$, where otherwise the Higgs sector
behaves mainly SM-like.

\section*{Acknowledgments}
The author thanks G.~Isidori, T.~Junk and S.~Heinemeyer
for useful discussions and the organizers of the
XXXVIth Rencontres de Moriond for the kind invitation and the pleasant
atmosphere enjoyed at Les Arcs.

\end{document}